\documentclass[twocolumn,showpacs,preprintnumbers,amsmath,amssymb,superscriptaddress,floatfix,nofootinbib]{revtex4}

\usepackage{graphicx}
\usepackage{amsmath}
\usepackage{booktabs}
\usepackage{amsfonts}
\usepackage{amssymb}

\begin{document}
\bibliographystyle{unsrt}
\title{Photoproduction of $f_0(980)$ and $f_0(1500)$ resonances off a proton target}
\date{\today}

\author{Hanyang Xing}
\affiliation{School of Physical Science and Technology, Southwest University, Chongqing 400715, China} \affiliation{Institute of Modern Physics, Chinese Academy of
Sciences, Lanzhou 730000, China}

\author{Chun-Sheng An}
\affiliation{School of Physical Science and Technology, Southwest University, Chongqing 400715, China}

\author{Ju-Jun Xie}
\affiliation{Institute of Modern Physics, Chinese Academy of
Sciences, Lanzhou 730000, China}

\author{Gang Li}~\email{gli@mail.qfnu.edu.cn}
\affiliation{School of Physics and Engineering, Qufu Normal University, Qufu 273165, China}

\begin{abstract}

We study the $\gamma p \to p f_0$ [$f_0(980)$ and $f_0(1500)$] reaction close to threshold within an effective Lagrangian approach. The production process is described by $s$-channel nucleon pole or $t$-channel $\rho$ and $\omega$ exchange. The $K^0 \bar{K}^0$ invariant mass distributions of the $\gamma p\to f_0(980)p \to K^0 \bar{K}^0$ and $\gamma p\to f_0(1500)p \to K^0 \bar{K}^0 p$ reactions are investigated, where the two kaons have been separated in $S$ wave decaying from $f_0(980)$ and $f_0(1500)$. It is shown that the $s$-channel process is favored for the production of $f_0(980)$, while for the $f_0(1500)$ production, the experimental measurements can be described quite well by the $t$-channel process. It is expected that the theoretical results can be tested by further experiments at CLAS.

\end{abstract}

\maketitle

\section{Introduction}

The study of the structure of low-lying scalar mesons is a topic of
high interest in hadronic physics and is attracting much
attention~\cite{Oset:2016lyh,Guo:2017jvc}. For the scalar meson
$f_0(980)$, it is now widely accepted that the simplest picture,
where it is described as an orbital excitation of quark-antiquark
pairs, is not compatible with the experimental observations on its
decay modes. Thus, the $f_0(980)$ is thought to be a molecule state
formed from the interaction of pseduoscalar
mesons~\cite{Weinstein:1982gc,Close:1992ay,Oller:1997ti,Oller:1998zr,Nieves:1999bx,Baru:2003qq,Pelaez:2003dy,Giacosa:2007bn},
and it couples strongly to the $K \bar{K}$
channel~\cite{Lee:2013mfa}, which is its dominant component. The
$f_0(1500)$, on the other hand, with a mass of $1504 \pm 6$ MeV and
a width of $109 \pm 7$ MeV~\cite{pdg2018}, is a candidate for having
much glueball content~\cite{Albaladejo:2008qa,Crede:2008vw}.
Photoproduction of these scalar resonances provides a unique place
to probe their nature.

On the experimental side, photoproduction of $f_0(980)$ meson on
protons was measured by CLAS collaboration in
Refs.~\cite{Battaglieri:2008ps,Battaglieri:2009aa} at the photon
energy region of $E_\gamma = 3.0 - 3.8$ GeV, where $f_0(980)$ was
detected via its decay in the $\pi^+\pi^-$ channel by performing a
partial wave analysis of the reaction $\gamma p \to p \pi^+\pi^-$.
However, the production rate of $f_0(980)$ is much smaller than the
one for the $\rho$ meson. Very recently, a partial wave analysis is
performed for the $\gamma p \to p K^+ K^-$ reaction by the CLAS
collaboration~\cite{Lombardo:2018gog}, where the production
amplitudes have been parametrized using a Regge-theory inspired
model. There were also pioneering
measurements~\cite{Fries:1978di,Barber:1981fj} for the
photoproduction of $K^+ K^-$ pairs. After that, there were several
theoretical calculations about the scalar mesons production in the
process of $\gamma p$ scattering. A combined analysis of $\pi \pi$
and $K\bar{K}$ photoproduction in $S$-wave is conducted in
Ref.~\cite{Ji:1997fb}, while the $f_0(980)$ and $a_0(980)$
photoproduction for photon energies close to the $K\bar{K}$
production threshold was studied in Ref.~\cite{Marco:1999nx} using
tools of chiral unitary approach. In Ref.~\cite{daSilva:2013yka},
within a model based on the Regge approach, a theoretical analysis
of the data on photoproduction of the $f_0(980)$ was done, where it
was shown that the radiative decay rate for $f_0(980) \to \gamma V$
is important in the theoretical predictions. In
Ref.~\cite{Tarasov:2013yma} the $\gamma p \to a_0(980) p$ and
$\gamma p \to f_0(980)p$ reactions were investigated with the main
aim for studying the possibility of observing $a_0(980)$-$f_0(980)$
mixing in these processes. In Ref.~\cite{Donnachie:2015jaa}, the
$a_0(980)$ and $f_0(980)$ photoproduction was investigated by
considering the Regge-cut effects which were fixed from $\pi^0$
photoproduction. With the Regge theory, the $f_0(1500)$
photoproduction was also studied in Ref.~\cite{daSilva:2011hy} at
$E_\gamma = 9$ GeV.

Recently, the reaction $\gamma p \to p X \to p K^0_S K^0_S$ was investigated by the CLAS Collaboration~\cite{Chandavar:2017lgs} with photon energies from $2.7-5.1$ GeV, where it was found that the angular distributions of the data suggest that most of the $K^0_S K^0_S$ decay is from scalar mesons in $S$ wave. In particular, a clear peak is seen at $1500$ MeV in the invariant mass spectra of $K^0_S K^0_S$, and the mass and width of this peak is consistent with that of the scalar meson $f_0(1500)$, while the enhancement close to $K^0_S K^0_S$ threshold is due to the $f_0(980)$ decay. In addition, there is no clear signals for contributions from the baryon resonances.

In the present work, based on the new measurements of CLAS
collaboration~\cite{Chandavar:2017lgs}, we reanalyze the $\gamma p
\to f_0(980) p \to K^0 \bar{K}^0 p$ and $\gamma p \to f_0(1500) p
\to K^0 \bar{K}^0 p$ reactions~\footnote{We take $|K^0> =
\frac{1}{\sqrt{2}}(|K^0_S> + |K^0_L>)$ and $|\bar{K}^0> =
\frac{1}{\sqrt{2}}(|K^0_S> - |K^0_L>)$, where we ignore the $CP$
violation.} within an effective Lagrangian method near threshold. As
in Refs.~\cite{daSilva:2013yka,Tarasov:2013yma} we consider the
contributions from $t$-channel $\rho^0$ and $\omega$ exchange. Since
the couplings of $f_0$ to $V\gamma$ channel is
scarce~\cite{pdg2018}, we take these results obtained in
Refs.~\cite{Nagahiro:2008mn,Nagahiro:2008bn}, where meson loops were
considered, and the $f_0(980)$ was taken as a dynamically generated
state. On the other hand, possible $s$-channel proton pole process,
which was not included in all these above theoretical calculations,
is also investigated in this work. It is shown that the new
measurements of Ref.~\cite{Chandavar:2017lgs} may indicate the
dominant $s$-channel contribution for the $f_0(980)$
photoproduction. In this respect, we show in this work how the CLAS
measurements could be used to determine the reaction mechanisms of
the photoproduction of these scalar mesons.

In the next section, we will give the formalism and ingredients in this work, then numerical results and discussions are given in Sec.~\ref{sec:results}. A short summary is given in the last section.

\section{Formalism and ingredients} \label{sec:formalism}

The effective Lagrangian method is widely used to calculate cross sections for different reactions in the resonance production region. In this section, we introduce theoretical formalism and ingredients to calculate the scalar mesons photoproduction off protons within the effective Lagrangian method.

\subsection{Interaction Lagrangian densities and scattering amplitudes}

\begin{figure}[htbp]
\begin{center}
\includegraphics[width=0.4\textwidth]{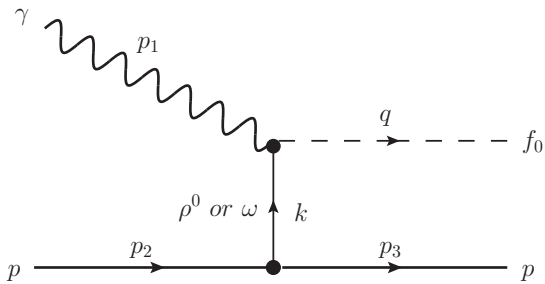}
\caption{Schematic diagram of reaction mechanism for $\gamma p \to f_0 p$ reaction with $t$-channel $\rho^0$ and $\omega$ exchange. The definition of the kinematical variables ($p_1$, $p_2$, $p_3$, $q$) used in the present calculation are also shown.} \label{fig:T2body}
\end{center}
\end{figure}

We first consider the basic $t$-channel tree level diagram for the $\gamma p \to p f_0$ [$f_0 \equiv f_0(980)$ or $f_0(1500)$] reaction as shown in Fig.~\ref{fig:T2body}. This includes the contributions from $\rho^0$ and $\omega$ meson exchange terms.

Following Ref.~\cite{Nagahiro:2008bn}, we can write down the amplitude for the $f_0 \to V \gamma$ deacy as
\begin{equation}
  T=-\frac{g_{f_0 V \gamma}}{m_{f_0}} (k \cdot p_1 g^{\mu \nu}-k^\mu p_1^\nu)\varepsilon_{V \mu}(k)\varepsilon_{\nu}(p_1),
\end{equation}
from where, we can obtain the partial decay width of the $f_0$ meson into a vector meson and a photon,
\begin{eqnarray}
 \Gamma_{f_0 \rightarrow V \gamma} &=& \frac{|\vec{k}|}{8\pi M^2_{f_0}} \sum\sum |T|^2 \nonumber \\
& =& \frac{g_{f_0 V \gamma}^2}{32\pi} \frac{(M_{f_0}^2-m_V^2)^3}{M_{f_0}^5}.
\end{eqnarray}

With masses ($M_{f_0(980)} = 990$ MeV, $M_{f_0(1500)} = 1504$ MeV, and $m_{\rho} = m_{\omega} = m_V = 780$ MeV), and the partial decay widths of the scalar $f_0(980)$ and $f_0(1500)$ mesons radiative decay into a vector meson and a photon as obtained in Refs.~\cite{Nagahiro:2008mn,Nagahiro:2008bn}, we obtain these coupling constants as list in Table~\ref{tab1}.

\begin{center}
\begin{table}[htbp]
\caption{Values of the coupling constants required for the estimation of the $\gamma p \to p f_0$ reaction.}
\begin{tabular}{|c|c|c|}
\hline \hline
   Decay channels  & Partial decay width $\Gamma_{f_0 \to V\gamma} (\mathrm{keV})$ & $g_{f_0 V\gamma}$ \\
  \hline
  $f_0(980) \to \rho \gamma$     & $7.3\pm 1.8$  & 0.12 \\
  $f_0(980) \to \omega \gamma$   & $6.6\pm 1.8$  & 0.11 \\   \hline
  $f_0(1500) \to \rho \gamma$    & $77\pm 8$     & 0.11 \\
  $f_0(1500) \to \omega \gamma$  & $79\pm 8$     & 0.12 \\
  \hline \hline
\end{tabular}
  \label{tab1}
\end{table}
  \end{center}

To compute the scattering amplitudes of the diagrams shown in Fig.~\ref{fig:T2body}, we need also the effective interactions for the $\rho NN$ and $\omega NN$ vertices. We take the interaction Lagrangian densities as used in Refs.~\cite{Machleidt:1987hj,Kochelev:2009xz}:
\begin{eqnarray}
\mathcal{L}_{\rho NN} &=& -g_{\rho NN}\bar{N}(\gamma^\mu-\frac{\kappa_{\rho}}{2 m_N}\sigma^{\mu \nu}\partial_\nu)\vec{\tau} \cdot \vec{\rho}_{\mu} N ,\\
\mathcal{L}_{\omega NN} &=& -g_{\omega NN}\bar{N}(\gamma^\mu-\frac{\kappa_{\omega}}{2 m_N}\sigma^{\mu \nu}\partial_\nu)\omega_{\mu} N.
\end{eqnarray}

We use the coupling constants $g_{\rho NN}=3.36$, $\kappa_\rho=6.1$, $g_{\omega NN}=15.85$ and $\kappa_\omega=0$ of Refs.~\cite{Xie:2014twa,Cheng:2016hxi}. Then we can write the $\rho NN$ and $\omega NN$ vertices as,
\begin{eqnarray}
  -\mathrm{i} t_{\rho NN} &=& \mathrm{i} g_{\rho NN}(\gamma^\mu+\mathrm{i} \frac{\kappa_{\rho}}{2 m_N}\sigma^{\mu \nu}q_\nu)\varepsilon_{\mu}(\rho), \\
  -\mathrm{i} t_{\omega NN} &=& \mathrm{i} g_{\omega NN}\gamma^\mu\varepsilon_{\mu}(\omega).
\end{eqnarray}

\subsection{$\gamma p \to p f_0$ scattering amplitudes}

With ingredients given above, we can easily obtain the $t$-channel $\gamma p \to f_0 p$ reaction invariant scattering amplitude:
\begin{equation}
\begin{split}
   \mathcal{M}_{V} & =-\bar{u}(p_3)\frac{g_{f_0 V \gamma}}{m_{f_0}}g_{VNN}(k \cdot p_1 g^{\nu \sigma}-k^\nu p_1^\sigma) G_{\mu \nu} \\
  & \times [\gamma_\mu+\frac{\kappa_{V}}{2m_p}(k_\mu-\not k \gamma_\mu)] F_1 u(p_2,s_p) \varepsilon_\sigma(p_1),
\end{split}
\end{equation}
where $G_{\mu \nu}$ is the Feynman propagator of $\rho$ or $\omega$ meson which has the following form:
\begin{equation}
G_{\mu \nu}=-\mathrm{i}\frac{g_{\mu \nu}-k_\mu k_\nu/m^2_{V}}{k^2-m^2_{V}}.
\end{equation}

Since hadrons are not point-like particles, the form factor of hadrons need to be taken into account~\cite{Cheng:2016hxi,Xie:2015wja}:
\begin{equation}
  F_1=(\frac{\Lambda^2_c-m^2_V}{\Lambda^2_c-t})^2,
\end{equation}
with $t = k^2$ and $\Lambda_c$ a free cut-off parameter.

\subsection{Differential cross section}

The differential cross section for the $\gamma p \to p f_0$ reaction by the exchanged $\rho^0$ and $\omega$ mesons can be expressed as
\begin{equation}\label{eq:22dc}
  \frac{d\sigma}{dt} =\frac{1}{16\pi s}\frac{m^2_p}{|\vec{p}_1|^2} \left(\frac{1}{4}\sum |\mathcal{M}|^2 \right ),
\end{equation}
where $s$ is the invariant mass square of the $\gamma p$ system, and $\vec{p}_1$ denotes the photon three momentum in the center of mass (c.m.) frame. The total invariant scattering amplitude ${\cal M}$ is given by
\begin{eqnarray}
&&\sum |\mathcal{M}|^2 = \sum|\mathcal{M}_\rho+\mathcal{M}_\omega|^2 \nonumber \\
 &=& \frac{1}{4m^2_p} \sum_{V_1, V_2=\rho, \omega}Tr[(\not{p_3}+m_3)\Gamma_{V_1}^\mu(\not{p_2}+m_2)\Gamma_{V_2}^\nu g_{\mu \nu}] \nonumber
\end{eqnarray}
with
\begin{eqnarray}
  \Gamma_V^\mu & =& \frac{g_{VNN}g_{{f_0}V\gamma}}{(t-m_V^2)M_{f_0}}[(1+\kappa_V) p_1\cdot k \gamma^\mu-(1+\kappa_V)\not{p_1} k^\mu  \nonumber \\
  & +&\frac{\kappa_V}{2M_N}p_1\cdot k (p_2^\mu+p_3^\mu)+\frac{\kappa_V}{2M_N}p_1 \cdot (p_2+p_3)k^\mu].
\end{eqnarray}

\begin{figure}[htp]
\begin{center}
  \includegraphics[scale=0.55]{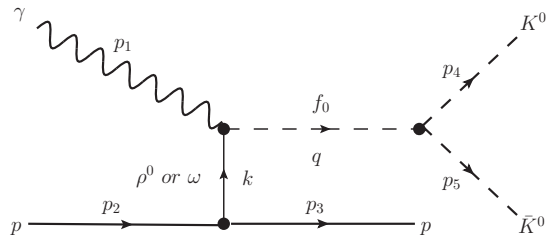}
  \caption{Feynman diagram for $t$-channel $\gamma p \to f_0 p \to K^0 \bar{K}^0 p$ reaction.}   \label{fig:t3body}
\end{center}
\end{figure}

On the other hand, we can generalize the two body process as in
Eq.~\eqref{eq:22dc} by considering the situation which allows the
$f_0 $ to decay into a $K^0$ and a $\bar{K}^0$ as shown in
Fig.~\ref{fig:t3body}. By working out the three-body phase space of
the $\gamma p \to f_0 p \to p K^0 \bar{K}^0 $ reaction, we find

\begin{equation}
\begin{split}
  \frac{d^2\sigma}{dM_{\rm inv}dt}  & =
  \frac{m^2_p}{32 \pi^2}\frac{M_{\rm inv}^2}{s|\vec{p}_1|^2} |\mathcal{M}|^2 \frac{ \Gamma_{f_0 \rightarrow K^0 \bar{K}^0}}{(M_{\rm inv}^2-M_{f_0}^2)^2+M_{f_0}^2\Gamma_{f_0}^2},
\end{split}
\end{equation}
where $\Gamma_{f_0}$ is the total decay width and we take
$\Gamma_{f_0} = 100$ MeV~\footnote{We take a relative large value
for the total decay width of the $f_0(980)$, which is favored by the
new CLAS measurements~\cite{Chandavar:2017lgs}.} and $109$ MeV for
$f_0(980)$ and $f_0(1500)$, respectively. $M_{\rm inv}$ represents
the invariant mass of $K^0 \bar{K}^0$. For $f_0(1500)$, $\Gamma_{f_0
\rightarrow K^0 \bar{K}^0}$ is given by
\begin{equation}
  \Gamma_{f_0 \rightarrow K^0 \bar{K}^0} = \Gamma^{\rm on}_{K^0 \bar{K}^0} \sqrt{\frac{M_{\rm inv}^2 - 4m^2_{K^0}}{M^2_{f_0} - 4 m^2_{K^0}}} \frac{M^2_{f_0}}{M_{\rm inv}^2},
\end{equation}
with $\Gamma^{\rm on}_{K^0\bar{K}^0} = 4.7$ MeV. While for the case of $f_0(980)$, we take
\begin{equation}
\Gamma_{f_0 \rightarrow K^0 \bar{K}^0} = \frac{g_{f_0 K \bar{K}}^2}{16 \pi} \frac{\sqrt{M^2_{\rm inv} - 4m^2_{K^0}}}{2M_{\rm inv}^2},
\end{equation}
with $g_{f_0(980)K\bar{K}} = 3860$ MeV as in Refs.~\cite{Nagahiro:2008bn,Oller:1998hw,Oller:2003vf}.

\section{Numerical results and discussion} \label{sec:results}

In this section we will show the numerical results for the $\gamma p \to p f_0$ reaction. We firstly show the theoretical results for the case of $f_0(1500)$ photoproduction.

\subsection{Invariant mass distributions for the $\gamma p \to p f_0(1500) \to p K^0 \bar{K}^0$ reaction}

We compare our theoretical calculations for the invariant $K^0 \bar{K}^0$ mass distributions as a function of $M_{\rm inv}$ with the recent CLAS data of Ref.~\cite{Chandavar:2017lgs}. The theoretical $d\sigma/{dM_{\rm inv}}$ is calculated by
\begin{equation}
 \frac{d\sigma}{dM_{\rm inv}} = \frac{ \int^{E^{\rm max}_{\gamma}}_{E^{\rm min}_{\gamma}}   dE_{\gamma} \int dt \frac{\mathrm{d}^2 \sigma}{d M_{\rm inv} dt}}{ \int^{E^{\rm max}_{\gamma}}_{E^{\rm min}_{\gamma}} dE_{\gamma}},
\end{equation}
with $E^{\rm max}_\gamma = 5.1$ GeV and $E^{\rm min}_{\gamma} = 2.7$ GeV, which are the photon energy region of Ref.~\cite{Chandavar:2017lgs}.

\begin{figure}[htbp]
\begin{center}
\includegraphics[width=0.4\textwidth]{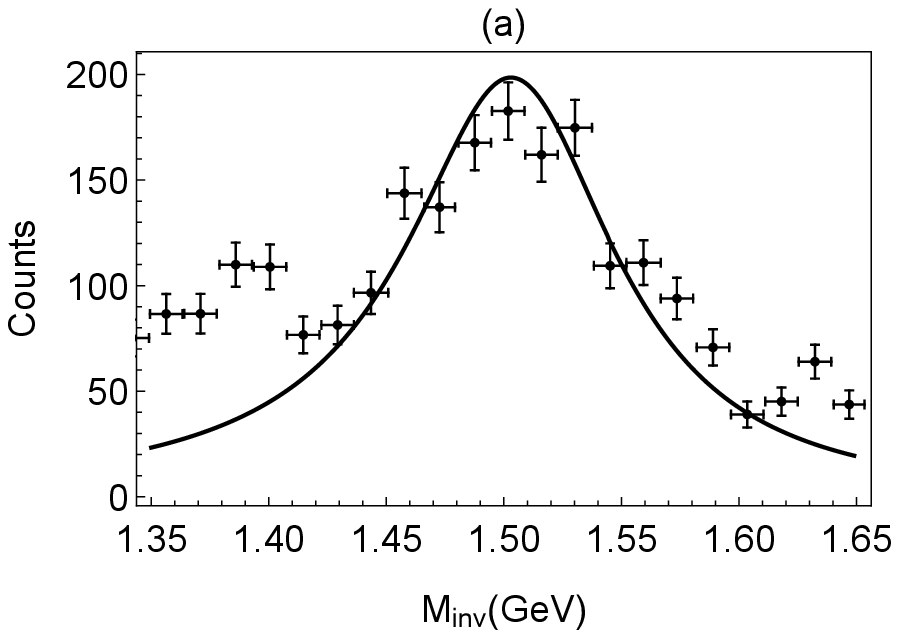}
\includegraphics[width=0.4\textwidth]{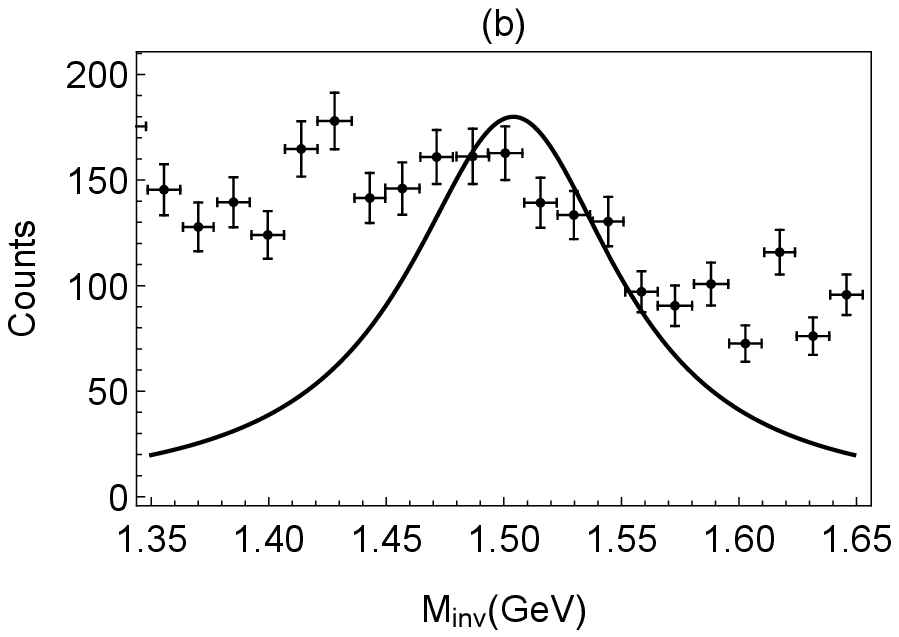}
\caption{Invariant mass distribution of $K^0 \bar{K}^0$ of $\gamma p \to p f_0(1500) \to p K^0 \bar{K}^0$ reaction for (a) $|t|<1.0~ \mathrm{GeV^2}$ and (b) $|t|>1.0~ \mathrm{GeV^2}$.}\label{fig:dsigdm1500}
\end{center}
\end{figure}

In Fig.~\ref{fig:dsigdm1500}, we show the theoretical results, ${c_1 d\sigma}/{dM_{\rm inv}}$, for the $K^0 \bar{K}^0$ invariant mass distributions for the $\gamma p \to p f_0(1500) \to p K^0\bar{K}^0$ reaction, comparing with the experimental measurements of Ref.~\cite{Chandavar:2017lgs}, where $c_1 = 2.2$ and $\Lambda_c = 1.7$ GeV has been adjusted to the strength of the experimental data reported by the CLAS Collaboration~\cite{Chandavar:2017lgs} at its peak around $M_{\rm inv}=1500$ MeV. One can see that, we can describe quite well the experimental measurements for the $\gamma p \to p f_0(1500) \to p K^0\bar{K}^0$ reaction by considering the $t$-channel $\rho^0$ and $\omega$ exchange, especially for the case of $|t|<1 ~ {\rm GeV}^2$. This may indicate that the $t$-channel process is dominant for the photoproduction of the $f_0(1500)$ resonance.

\subsection{Invariant mass distributions for the $\gamma p \to p f_0(980) \to p K^0 \bar{K}^0$ reaction}

\begin{figure}[htbp]
\begin{center}
\includegraphics[width=0.4\textwidth]{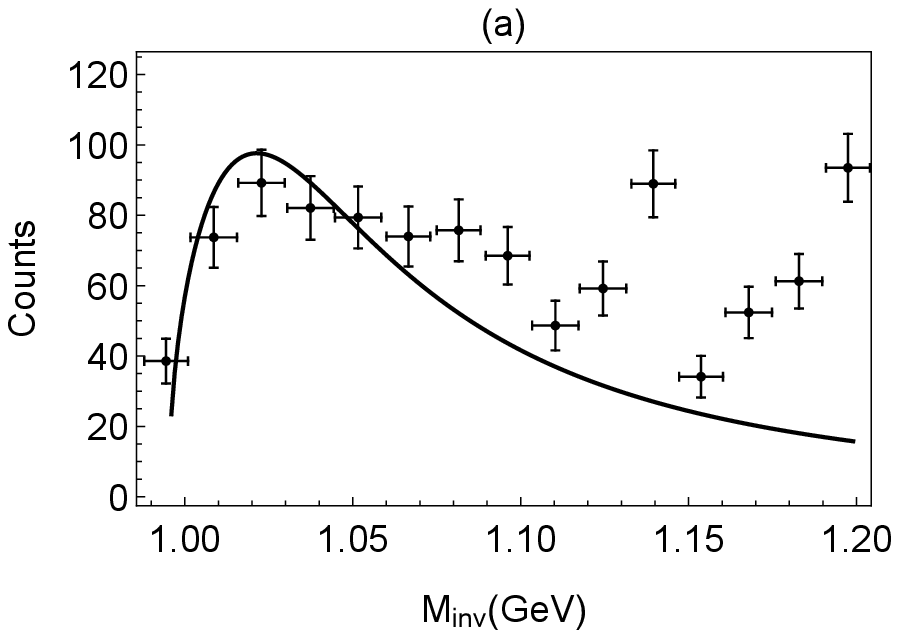}
\includegraphics[width=0.4\textwidth]{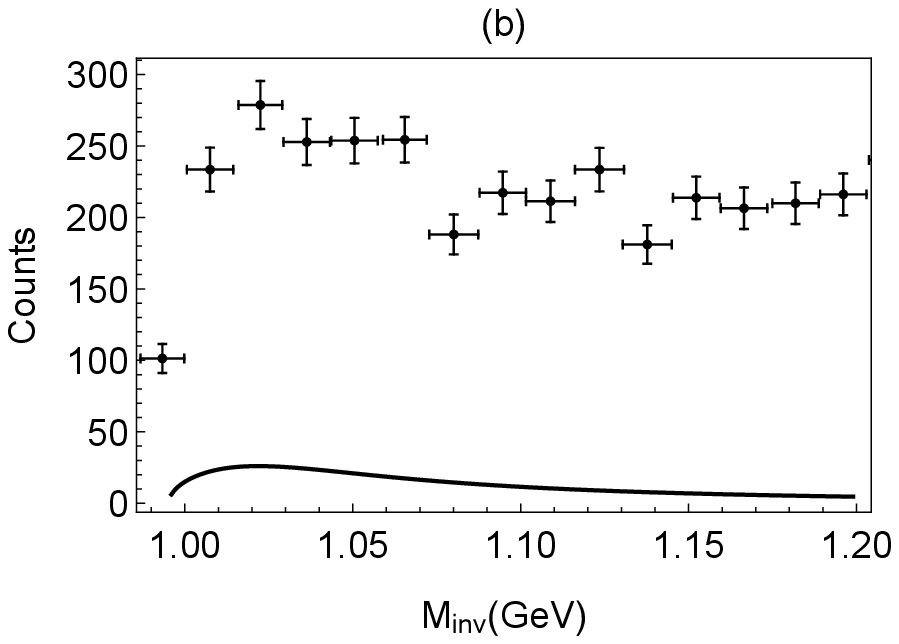}
  \caption{Invariant mass distribution of $K^0 \bar{K}^0$ of $\gamma p \to p f_0(980) \to p K^0 \bar{K}^0$ reaction for (a) $|t|<1.0~ \mathrm{GeV^2}$ and (b) $|t|>1.0~ \mathrm{GeV^2}$. The theoretical results are obtained by considering only $t$-channel $\rho ^0$ and $\omega$ exchange. }\label{fig:dsigdm980-tchannel}
\end{center}
\end{figure}

We first present the theoretical results for the $\gamma p \to p
f_0(980) \to p K^0 \bar{K}^0$ reaction by including the $t$-channel
$\rho^0$ and $\omega$ exchange. The numerical results of the $K^0
\bar{K}^0$ invariant mass distributions obtained with $c_1=0.9$ and
$\Lambda_c = 1.07$ GeV, are shown in
Fig.~\ref{fig:dsigdm980-tchannel}. The peak of the $K^0\bar{K}^0$
invariant mass distributions is around $1020$ MeV, very close to the
mass threshold ($995$ MeV) of $K^0\bar{K}^0$. One can see that the
model cannot describe simultaneously both the experimental data for
$|t|<1~{\rm GeV}^2$ and $|t|>1$ GeV$^2$. At $M_{\rm inv} = 1020$ MeV
and $E_\gamma = 3.9$ GeV, the values of $t$ is $-5.36~{\rm
GeV}^2<t<-0.02 ~{\rm GeV}^2$,~\footnote{The values of $t$ for the
production of $f_0(1500)$ is $-3.89~{\rm GeV}^2<t<-0.14~{\rm
GeV}^2$.} from where we find that the phase space for $|t|>1~{\rm
GeV}^2$ is more than four times larger than the case of $|t|<1 ~{\rm
GeV}^2$. However, the $t$-channel form factor $F_1 =
(\frac{\Lambda^2_c-m^2_V}{\Lambda^2_c-t})^2$ with $\Lambda_c \sim
1.07$ GeV will contribute a suppression with factor about 14 for the
case of $|t|>1~{\rm GeV}^2$ than that of $|t|<1 ~{\rm GeV}^2$.
Hence, it is expected that, with the values of $c_1=0.9$ and
$\Lambda_c = 1.07$ GeV, the results for $|t|>1~{\rm GeV}^2$ should
be much smaller than the ones for $|t|<1 ~ {\rm GeV}^2$. But, the
experimental data of Ref.~\cite{Chandavar:2017lgs} tell us that the
values for $|t|>1~{\rm GeV}^2$ are even larger than those for
$|t|<1~ {\rm GeV}^2$. This may indicate that the $t$-channel
exchange mechanism is not enough to explain the experimental
measurements of CLAS Collaboration~\cite{Chandavar:2017lgs}.

We have also performed calculations for the $\gamma p \to p f_0(980)
\to p K^0 \bar{K}^0$ reaction with different values of $c_1$ and
$\Lambda_c$. It turns out that we can also reproduce the
experimental measurements with $c_1 = 0.012$ and a large $\Lambda_c
= 5$ GeV. Thus, the inclusion of other reaction mechanism is needed
to achieve a good description of the new CLAS experimental
measurements.

\begin{figure}
\begin{center}
  \includegraphics[width=0.4\textwidth]{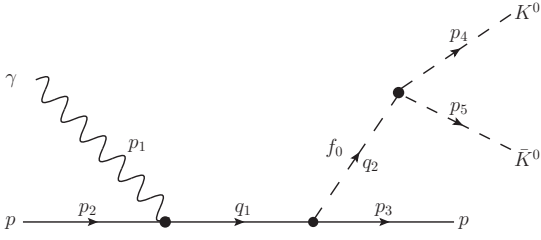}
  \caption{Feynman diagram for $s$-channel $\gamma p\rightarrow f_0(980) p \rightarrow K^0 \bar{K}^0 p$ reaction.}
  \label{fig:s3body}
\end{center}
\end{figure}

Next, we study another kind of reaction mechanism for $\gamma p \to p f_0(980) \to p K^0 \bar{K}^0$ reaction, which is depicted in Fig.~\ref{fig:s3body}, where we have considered the contribution from the $s$-channel nucleon pole term. To compute the contribution of this term, the interaction Lagrangian densities for $\gamma pp$ and $f_0(980) pp$ vertexes are needed. We take them as used in Refs.~\cite{Tarasov:2013yma,Xie:2010yk}:
\begin{eqnarray}
\mathcal{L}_{\gamma pp} &=& -e\bar{p} [A\!\!\!/-\frac{\kappa_{p}}{2 m_N}\sigma^{\mu \nu}(\partial_\nu A_\mu)]p , \\
\mathcal{L}_{f_0(980) pp} &=& g_{f_0 pp}\bar{p} p f_0 ,
\end{eqnarray}
where $\kappa_p=1.5$.

Then one can easily write down the corresponding amplitude for $s$-channel nucleon pole term as,
\begin{equation}
\begin{split}
\mathcal{M}_{s} & = g_{f_0pp} F_2 \bar{u}(p_3)\frac{{q_1}\!\!\!\!\!/+m_p}{q_1^2-m_p^2} \\
& \times (\gamma^\mu - \Gamma_{c}^{\mu} - \frac{\kappa_p}{2m_p}\gamma^\mu {p_1}\!\!\!\!\!/) u(p_2) \varepsilon_\mu(p_1),
\end{split}
\end{equation}
with
\begin{equation}
F_2=\frac{\Lambda^4_s}{\Lambda^4_s+(q_1^2-m_p^2)^2}
\end{equation}
and
\begin{equation}
\Gamma_c^\mu = -\frac{p_1\!\!\!\!\!/}{p_1 \cdot p_2}p_2^\mu,
\end{equation}
which is obtained from a contact term and for keeping the scattering amplitude $\mathcal{M}_{s}$ gauge invariant~\cite{Xie:2010yk,Haberzettl:2006bn}.

\begin{figure}[htbp]
\begin{center}
\includegraphics[width=0.4\textwidth]{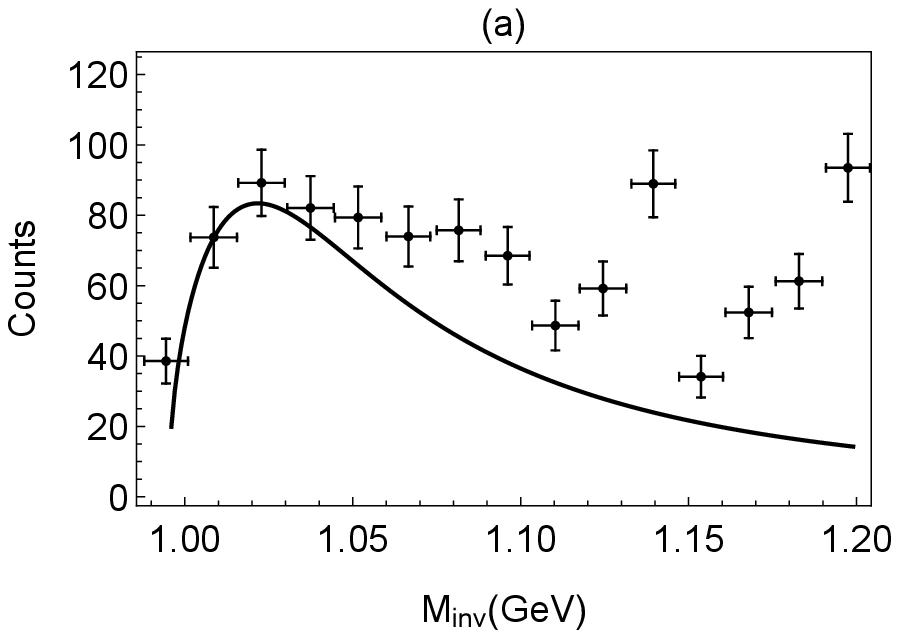}
\includegraphics[width=0.4\textwidth]{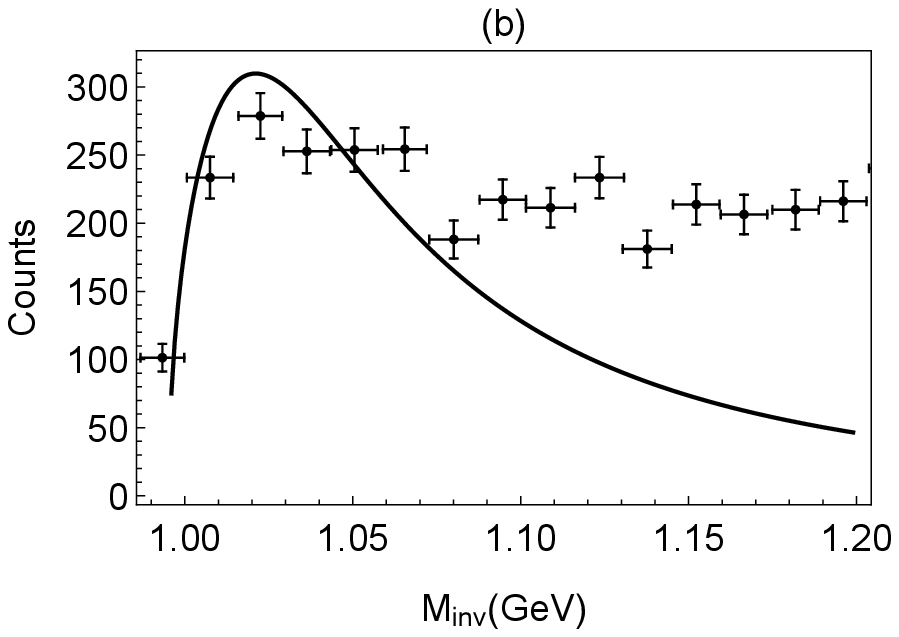}
  \caption{Invariant mass distribution of $K^0 \bar{K}^0$ of $\gamma p \to p f_0(980) \to p K^0 \bar{K}^0$ reaction for (a) $|t|<1.0~ \mathrm{GeV^2}$ and (b) $|t|>1.0~ \mathrm{GeV^2}$. The theoretical results are obtained with contribution from the $s$-channel nucleon pole.}\label{fig:dsigdm980-schannel}
\end{center}
\end{figure}

The theoretical results of $K^0 \bar{K}^0$ invariant mass
distributions of the $\gamma p\rightarrow f_0(980) p \rightarrow K^0
\bar{K}^0 p$ reaction with the contribution from $s$-channel nucleon
pole are shown in Fig.~\ref{fig:dsigdm980-schannel}, from where we
can see that we can explain the experimental measurements for both
$|t|<1~{\rm GeV}^2$ and $|t|>1 ~{\rm GeV}^2$ cases quite well, since
there is no so strong $t$ dependence factor $F_1$ in the $s$-channel
process. The theoretical numerical results shown in
Fig.~\ref{fig:dsigdm980-schannel} are obtained with $c_1 g_{f_0 pp}
=7.5$ and $\Lambda_s = 1.1$ GeV.

One might think that the inclusion of higher nucleon excitations might improve the situation, since they have large mass and will give large contributions. However, at one certain photon energy $E_\gamma$, the propagator of the $s$-channel process is then just a constant. The estimation of the $K^0 \bar{K}^0$ invariant mass distributions in our
model is only sensitive to the production rate of the $f_0(980)$,
and the nucleon pole term is sufficient for this purpose. By
neglecting contribution from other $N^*$ resonances, we can present a more general picture
of the $s$-channel $f_0(980)$ production processes, though our
results are more general than this would suggest.

\begin{figure}[htbp]
\begin{center}
\includegraphics[width=0.4\textwidth]{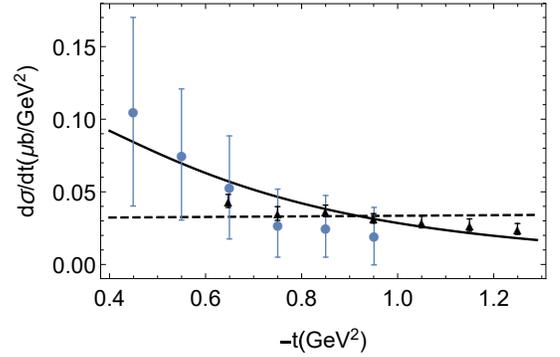}
\caption{Differential cross sections $d\sigma/dt$ of $\gamma p \to p
f_0(980)$ reaction compared with the CLAS data from
Ref.~\cite{Battaglieri:2008ps} (dot) and
Ref.~\cite{Lombardo:2018gog} (triangle).}\label{fig:dsigdt980}
\end{center}
\end{figure}

On the other hand, we calculate $d\sigma/dt$ for $\gamma p \to p
f_0(980)$ reaction with the above two different reaction mechanisms
at the photon energy $E_\gamma = 3.4$ GeV. The numerical
results~\footnote{The results for the $s$-channel process are
obtained with $g_{f_0 pp} = 4.32$.} are shown in
Fig.~\ref{fig:dsigdt980}, comparing with the experimental data taken
from Refs.~\cite{Battaglieri:2008ps,Lombardo:2018gog}. The solid and
dashed lines represent the result from $t$- and $s$-channel process,
respectively. One can see that both $t$-channel mechanism and
$s$-channel process can describe fairly well the current
experimental data. However, the line shapes of these two different
reaction mechanisms are sizably different. The slope of the results
for the $s$-channel process is more flat than the case of
$t$-channel $\rho^0$ and $\omega$ exchange. We hope that this
feature may be used to determine the reaction mechanism of $\gamma p
\to p f_0(980)$ reaction.

\section{Summary}

In this work, we have studied the $\gamma p \to p f_0 [f_0(980), f_0(1500)] \to p K^0\bar{K}^0$
reactions near threshold within an effective Lagrangian
approach. The $K^0 \bar{K}^0$ invariant mass distributions are evaluated, where the two kaons have been separated in $S$ wave decaying from the scalar mesons $f_0(980)$ and $f_0(1500)$. It is shown that the $t$-channel $\rho^0$ and $\omega$ exchange processes can describe the experimental data on the $\gamma p \to p f_0(1500) \to p K^0\bar{K}^0$
reaction, while the $s$-channel process is favored for the $\gamma p \to p f_0(980) \to p K^0\bar{K}^0$
reaction, since the $t$-channel mechanism for the $f_0(980)$ photoproduction fails to reproduce the experimental measurements. Furthermore, it is found
that the theoretical numerical results for the $\gamma p \to p f_0(980)$ differential cross section, $d\sigma /dt$, of the two different reaction mechanisms are sizeably different. It is expected that the theoretical results can be tested by further experimental measurements at CLAS~\cite{Chandavar:2017lgs}.

Finally, we would like to stress that, thanks to the
important role played by the non $t$-channel process in the
$\gamma p \to p f_0(980)$ reaction, accurate data for this reaction
can be used to improve our knowledge about the reaction mechanism of this reaction and also the nature of $f_0(980)$. This work
constitutes a first step in this direction.

\section*{Acknowledgements}

This work is partly supported by the National Natural Science Foundation of China under Grants No. 11475227, 11522539, 11675131, 11675091, and 11735003, the fundamental Research Funds for the Central Universities and the Youth Innovation Promotion Association CAS No. 2016367.


\begin{thebibliography}{99}

\bibitem{Oset:2016lyh}
  E.~Oset {\it et al.},
  Int.\ J.\ Mod.\ Phys.\ E {\bf 25}, 1630001 (2016).

\bibitem{Guo:2017jvc}
  F.~K.~Guo, C.~Hanhart, U.~G.~Meißner, Q.~Wang, Q.~Zhao and B.~S.~Zou,
  Rev.\ Mod.\ Phys.\  {\bf 90}, 015004 (2018).

\bibitem{Weinstein:1982gc}
  J.~D.~Weinstein and N.~Isgur,
  Phys.\ Rev.\ Lett.\  {\bf 48}, 659 (1982).

\bibitem{Close:1992ay}
  F.~E.~Close, N.~Isgur and S.~Kumano,
  Nucl.\ Phys.\ B {\bf 389}, 513 (1993).

\bibitem{Oller:1997ti}
  J.~A.~Oller and E.~Oset,
  Nucl.\ Phys.\ A {\bf 620}, 438 (1997)
  Erratum: [Nucl.\ Phys.\ A {\bf 652}, 407 (1999)].

\bibitem{Oller:1998zr}
  J.~A.~Oller and E.~Oset,
  Phys.\ Rev.\ D {\bf 60}, 074023 (1999).

\bibitem{Nieves:1999bx}
  J.~Nieves and E.~Ruiz Arriola,
  Nucl.\ Phys.\ A {\bf 679}, 57 (2000).

\bibitem{Baru:2003qq}
  V.~Baru, J.~Haidenbauer, C.~Hanhart, Y.~Kalashnikova and A.~E.~Kudryavtsev,
  Phys.\ Lett.\ B {\bf 586}, 53 (2004).

\bibitem{Pelaez:2003dy}
  J.~R.~Pelaez,
  Phys.\ Rev.\ Lett.\  {\bf 92}, 102001 (2004).

\bibitem{Giacosa:2007bn}
  F.~Giacosa and G.~Pagliara,
  Phys.\ Rev.\ C {\bf 76}, 065204 (2007).

\bibitem{Lee:2013mfa}
  H.~J.~Lee, N.~I.~Kochelev and Y.~Oh,
  Phys.\ Rev.\ D {\bf 87}, 117901 (2013).

\bibitem{pdg2018}
M. Tanabashi {\it et al.} (Particle Data Group), Phys. Rev. D {\bf 98}, 030001 (2018).

\bibitem{Albaladejo:2008qa}
  M.~Albaladejo and J.~A.~Oller,
  Phys.\ Rev.\ Lett.\  {\bf 101}, 252002 (2008).

\bibitem{Crede:2008vw}
  V.~Crede and C.~A.~Meyer,
  Prog.\ Part.\ Nucl.\ Phys.\  {\bf 63}, 74 (2009).


\bibitem{Battaglieri:2008ps}
  M.~Battaglieri {\it et al.} [CLAS Collaboration],
  Phys.\ Rev.\ Lett.\  {\bf 102}, 102001 (2009).

\bibitem{Battaglieri:2009aa}
  M.~Battaglieri {\it et al.} [CLAS Collaboration],
  Phys.\ Rev.\ D {\bf 80}, 072005 (2009).

\bibitem{Lombardo:2018gog}
  S.~Lombardo {\it et al.} [CLAS Collaboration],
  arXiv:1808.01918 [hep-ex].

\bibitem{Fries:1978di}
  D.~C.~Fries {\it et al.},
  Nucl.\ Phys.\ B {\bf 143}, 408 (1978).

\bibitem{Barber:1981fj}
  D.~P.~Barber {\it et al.},
  Z.\ Phys.\ C {\bf 12}, 1 (1982).

\bibitem{Ji:1997fb}
  C.~R.~Ji, R.~Kaminski, L.~Lesniak, A.~Szczepaniak and R.~Williams,
  Phys.\ Rev.\ C {\bf 58}, 1205 (1998).

\bibitem{Marco:1999nx}
  E.~Marco, E.~Oset and H.~Toki,
  Phys.\ Rev.\ C {\bf 60}, 015202 (1999).

\bibitem{daSilva:2013yka}
  M.~L.~L.~da Silva and M.~V.~T.~Machado,
  Phys.\ Rev.\ C {\bf 87}, 065201 (2013).

\bibitem{Tarasov:2013yma}
  V.~E.~Tarasov, W.~J.~Briscoe, W.~Gradl, A.~E.~Kudryavtsev and I.~I.~Strakovsky,
  Phys.\ Rev.\ C {\bf 88}, 035207 (2013).

\bibitem{Donnachie:2015jaa}
  A.~Donnachie and Y.~S.~Kalashnikova,
  Phys.\ Rev.\ C {\bf 93}, 025203 (2016).

\bibitem{daSilva:2011hy}
  M.~L.~L.~da Silva and M.~V.~T.~Machado,
  Phys.\ Rev.\ C {\bf 86}, 015209 (2012).

\bibitem{Chandavar:2017lgs}
  S.~Chandavar {\it et al.} [CLAS Collaboration],
  Phys.\ Rev.\ C {\bf 97}, 025203 (2018).

\bibitem{Nagahiro:2008mn}
  H.~Nagahiro, L.~Roca and E.~Oset,
  Eur.\ Phys.\ J.\ A {\bf 36}, 73 (2008).

\bibitem{Nagahiro:2008bn}
  H.~Nagahiro, L.~Roca, E.~Oset and B.~S.~Zou,
  Phys.\ Rev.\ D {\bf 78}, 014012 (2008).

\bibitem{Machleidt:1987hj}
  R.~Machleidt, K.~Holinde and C.~Elster,
  Phys.\ Rept.\  {\bf 149}, 1 (1987).

\bibitem{Kochelev:2009xz}
  N.~I.~Kochelev, M.~Battaglieri and R.~De Vita,
  Phys.\ Rev.\ C {\bf 80}, 025201 (2009).

\bibitem{Xie:2014twa}
  J.~J.~Xie and E.~Oset,
  Eur.\ Phys.\ J.\ A {\bf 51}, 111 (2015).

\bibitem{Cheng:2016hxi}
  C.~Cheng, J.~J.~Xie and X.~Cao,
  Commun.\ Theor.\ Phys.\  {\bf 66}, 675 (2016).


\bibitem{Xie:2015wja}
  J.~J.~Xie,
  Phys.\ Rev.\ C {\bf 92}, 065203 (2015).

\bibitem{Oller:1998hw}
  J.~A.~Oller, E.~Oset and J.~R.~Pelaez,
  Phys.\ Rev.\ D {\bf 59}, 074001 (1999)
  Erratum: [Phys.\ Rev.\ D {\bf 60}, 099906 (1999)]
  Erratum: [Phys.\ Rev.\ D {\bf 75}, 099903 (2007)].

\bibitem{Oller:2003vf}
  J.~A.~Oller,
  Nucl.\ Phys.\ A {\bf 727}, 353 (2003).

\bibitem{Xie:2010yk}
  J.~J.~Xie and J.~Nieves,
  Phys.\ Rev.\ C {\bf 82}, 045205 (2010).

\bibitem{Haberzettl:2006bn}
  H.~Haberzettl, K.~Nakayama and S.~Krewald,
  Phys.\ Rev.\ C {\bf 74}, 045202 (2006).
\end{thebibliography}
\end{document}